\documentstyle [12pt]{article}

\newcommand{\upa}{\uparrow}
\newcommand{\dna}{\downarrow}

\begin{document}

\newpage
\begin{center}
\Large
The Persistance of Memory: Surreal Trajectories in Bohm's Theory
\normalsize
\vspace{1in} 
Jeffrey A. Barrett
\end{center}
\vspace{1in}

\begin{abstract}
In this paper I describe the history of the surreal
trajectories problem and argue that in fact it is not 
a problem for Bohm's theory.  More specifically, I argue
that one can take the particle trajectories
predicted by Bohm's theory to be the actual trajectories
that particles follow and that there is no reason
to suppose that good particle detectors are somehow
fooled in the context of the surreal trajectories
experiments.  Rather than showing that Bohm's theory
predicts the wrong particle trajectories or that it
somehow prevents one from making reliable measurements,
such experiments ultimately reveal the special role
played by position and the
fundamental incompatibility between Bohm's theory and
the relativity.\footnote{This paper is an extension
of the discussion of surreal trajectories in Barrett
(1999, 127--40).  That section of the book was based on talk I gave
at the Quantum Mechanics Workshop at the University of Pittsburgh
in the Spring of 1997.  And that talk owed much to conversations
with Peter Lewis, David Albert, and Rob Clifton.}

\end{abstract}

\section{}

Bohm's theory\footnote{This is the theory
described by David Bohm in 1952.  Bohm's most complete
description of his theory is found in Bohm and Hiley (1993).}
has become increasingly popular as a nonrelativistic
solution to the quantum measurement problem.
It makes the same empirical predictions for the statistical distribution
of particle configurations as the standard von Neumann-Dirac
collapse formulation of quantum mechanics whenever the latter makes
unambiguous predictions.  Bohm's theory also treats measuring devices
exactly the same way it treats other physical systems.  The
quantum-mechanical state of a system
always evolves in the usual linear, deterministic way, so one
does not encounter the problems that arise in collapse formulations
of quantum mechanics when one tries to stipulate the conditions under which a
collapse occurs.  And Bohm's theory does not require one to postulate
branching worlds or disembodied minds or any of the other extravagant
assumptions that often
accompany no-collapse formulations of quantum mechanics. 

While Bohm's theory avoids many of the problems associated with other formulations
of quantum mechanics, it does have its own problems.  One problem,
it has been argued, is that the particle trajectories it predicts
are not the real particle trajectories.  This is the surreal
trajectories problem.  If Bohm's theory does in fact make false predictions
concerning particle trajectories, then this is presumably a serious problem.  I
will argue, however, that there is no reason to suppose that Bohm's theory
makes false predictions concerning the trajectories of particles.  Indeed,
I will argue that a good position measuring device need never
be mistaken concerning the actual position of a particle {\em at the
moment that the particle's position is in fact recorded}.

While surreal trajectories are not a problem for Bohm's theory,
the way that it accounts for the results of the surreal
trajectories experiments reveals the sense in which it
is fundamentally incompatible with relativity, and this is a problem.

\section{}

On Bohm's theory the quantum-mechanical
state $\psi$ evolves in the usual linear, deterministic way, but
one supposes that every particle always has a determinate position
and follows a continuous, determinsitic trajectory.  The motion of
a particular particle typically depends on the evolution of $\psi$ and the
positions of other (perhaps distant) particles.   The particle
motion is described by an auxiliary dynamics,
a dynamics that supplements the usual linear quantum dynamics.  In its
simplest form, what one might call the {\em minimal version\/}
(the version of the theory described
by Bell 1987, 127), Bohm's theory is characterized
by the following basic principles:

1.  State Description: The complete physical state at a time is given by the wave function $\psi$
and the determinate particle configuration $Q$.

2.  Wave Dynamics: The time evolution of the wave function is given by
the usual linear dynamics.  In
the simplest case, this is just Schr\"odinger's equation
\begin{equation}
i \hbar \frac{\partial \psi}{\partial t} =  \hat{H} \psi
\end{equation}
More generally, one uses the form of the linear dynamics appropriate to one's application (as in
the spin examples discussed below).

3.  Particle Dynamics: The particles move according to
\begin{equation}
\frac{d Q_k}{dt} = \frac{1}{m_k} \frac{\mbox{Im}(\psi^* \nabla_k \psi)}{\psi^* \psi}
\mbox{ evaluated at }Q
\end{equation}
where $m_k$ is the mass of particle $k$ and $Q$ is the current particle
configuration.

4.  Distribution Postulate: There is a time $t_0$ when the epistemic probability density for the
configuration $Q$ is given by $\rho(Q, t_0)= |\psi(Q, t_0)|^2$.

If there are $N$ particles, then $\psi$ is a function in $3N$-dimensional configuration space
(three dimensions for the position of each particle), and the current particle
configuration $Q$ is represented by a single point in configuration space (in configuration space
a single point gives the position of every particle).  Again, each particle moves in a way that
depends on its position, the evolution of the wave function, and the
positions of the other particles.

Concerning how one should think of the role of the wave function in Bohm's theory,
John Bell once said that ``{\em no one can understand this theory until he is willing to think of $\psi$ as a real
objective field rather than just a `probability amplitude.'  Even though it propagates not in 
$3$-space but in $3N$-space}'' (1987, 128).  While the ontology suggested
by Bell here is at best puzzling, the practical idea behind it is a good
one: The best way to picture what the particle dynamics does is to picture
the point representing the $N$-particle configuration being carried along
by the probability currents generated by the linear evolution
of the wave function $\psi$ in configuration space.  Once one has
this picture firmly in mind one will understand how Bohm's theory accounts
for quantum-mechanical correlations in the context of the surreal-trajectory
experiments and the sense in which the theory
is fundamentally incompatible with relativity.

Since the total particle configuration can be thought of as being pushed around by the probability
current in configuration space, the probability of the particle configuration being found in a
particular region of configuration space changes as the integral of $|\psi|^2$ over that region
changes.  More specifically, the continuity equation
\begin{equation}
\frac{\partial \rho}{\partial t} + \mbox{div}(\rho v^\psi) = 0
\end{equation}
is satisfied by the probability density $\rho=|\psi|^2$.  And this means that if the
epistemic probability density for the particle configuration is ever $|\psi|^2$, then it will always be
$|\psi|^2$, unless one makes an observation. That is,
if one starts with an epistemic probability density of $\rho(t_0)=|\psi(t_0)|^2$, then,
given the dynamics, one should update this probability density at time $t$ so
that $\rho(t)=|\psi(t)|^2$.  And if one makes an observation, then the
epistemic probability density will be given by the system's
{\em effective\/} wave function, the component (in the configuration space representation) of the total
wave function that is in fact responsible for the post-measurement time
evolution of the system's configuration.  The
upshot is that if the distribution postulate is ever
satisfied, then the most that one can learn from a measurement is the wave packet that the current
particle configuration is associated with and the epistemic probability distribution
for the actual configuration over this packet.\footnote{See D\"urr, D., S. Goldstein, and N. Zangh\'i
(1993) for a discussion of the equivariance of the statistical distribution $\rho$ and the notion of
the effective wave function in Bohm's theory.}  This is why Bohm's theory
makes the same statistical predictions for particle configurations
as the standard collapse formulation of quantum mechanics.

While it makes the same statistical predictions as the standard
formulation of quantum mechanics, Bohm's theory is deterministic.
More specifically, given the energy properties of a
simple closed system, the complete physical state at any time (the wave
function and the particle configuration) fully determines the physical state at all other
times.\footnote{We will say that a closed system is simple if the Hamiltonian
is bounded and if the particle configuration always has positive wave function
support.}.  It follows that, given a particular evolution of the wave function,
possible trajectories for the configuration of a system
can never cross at a time in configuration space.  And this feature of Bohm's theory
will prove important later.

Another feature of Bohm's theory that will prove imporant later is the
special role played by position in accounting for our determinate
measurement results.  In order for Bohm's
theory to explain why we get the determinate measurement records that we
do (which is presumably a precondition for it counting as a
solution to the measurement problem), one must suppose, as
a basic interpretational principle, that, given the usual
quantum mechanical state, making particle positions determinate
provides determinate measurement records.  Since particle
positions are always determinate on Bohm's theory, this would
guarantee determinate measurement records.  And, at least
on the minimal version of Bohm's theory, position is the
only determinate, noncontextual property that could serve
to provide determinate measurement records.\footnote{There
is a sense in which one might say that some dynamical
properties, like momentum, are also noncontextual in Bohm's theory,
but, as we will see, the noncontextual momentum
is not the {\em measured\/} momentum.}

The distinction between noncontextual and contextual properties
deserves some explanation.  Whether a system is found to have a particular contextual
property or not typically depends on how one measures the property:
one might get the result ``Yes'' if the contextual property is measured
one way and ``No'' if it is measured
another.  Consequently, contextual properties are 
not intrinsic properties of the system
to which they are typically ascribed.  One might say that  
contextual properties serve to prop up our talk of those
properties that we are used to talking about but which arguably should
not count as properties at all in Bohm's theory.  While the language of contextual properties provides
a convenient (but often misleading!) way of comparing the predictions
of Bohm's theory with the predictions of other physical theories, the
predictions of Bohm's theory are always ultmately just predictions
about the evolution of the wavefunction and the positions of the particles relative
to the wavefunction.  

The upshot of all this is just that position relative to the wave function,
or more precisely configuration relative to the wave function, is
ultimately the only property that one can appeal to in the minimal version
of Bohm's theory to explain how it is that we end up with the determinate measurement records
we do.  And this means that {\em for an interaction to count as a measurement,
it must produce a record in terms of the position of
something\/}---it must correlate the position of
something with some aspect of the quantum-mechanical state
of the system being measured.  So in order to explain our
determinate measurement records on Bohm's theory one must suppose that all
measurement records are ultimately position records, records represented
in the relationship between the particle configuration and the wave function.  And
since Bohm's theory predicts the right quantum
statistics for particle positions relative to the wave function,
it predicts the right quantum statistics for our measurement
records.\footnote{The point here is that making the right statistical predictions
concerning {\em particle configurations\/} is not necessarily a
sufficient condition for making the right statistical predictions for
our {\em measurement records}.  One needs to make an extra assumption about the
relationship between particle configurations and measurement records.}

\section{}

In their 1992 paper Englert, Scully, S\"ussman, and Walther (ESSW) argued
that the trajectories predicted by Bohm's theory are not the real trajectories followed by particles, but
rather are ``surreal''.  The worry is that the observed trajectories of
particles are not the trajectories that the particles actually follow in
Bohm's theory.  And if our observations are reliable and if Bohm's theory
predicts the wrong particle trajectories, then this is presumably
a problem for the theory.

ESSW describe the surreal trajectories problem in the
context of a two-path, delayed-choice interference experiment.
John Bell (1980, reprinted in 1987) was perhaps the first
to consider such an experiment in the context of Bohm's theory.

Consider an experiment where a spin-$1/2$ particle $P$ starts at region $S$ in a $z$-spin up
eigenstate, has its wave packet split into an $x$-spin up component that travels from $A$ to $A'$
and an $x$-spin down component that travels from $B$ to $B'$.\footnote{Since the standard line
is that position is the only observable physical quantity in Bohm's theory, this does not mean that
$P$ has a determinate $z$-spin; rather, it is just a description of the spin index associated with
$P$'s effective wave function.}

\vspace{5mm}
[Figure 1: Crossing-Paths Experiment]
\vspace{5mm}

The wave function evolves as follows:

\vspace{5mm}

\noindent
Initial state:
\begin{equation}
|\upa_z\rangle_P|S\rangle_P = 1/\sqrt{2} (|\upa_x\rangle_P +
|\dna_x\rangle_P)|S\rangle_P
\end{equation}

\noindent
After the initial wave packet splits:
\begin{equation}
1/\sqrt{2} (|\upa_x\rangle_P|A\rangle_P +
|\dna_x\rangle_P|B\rangle_P)
\end{equation}

\noindent
Final state:
\begin{equation}
1/\sqrt{2} (|\upa_x\rangle_P|A'\rangle_P + |\dna_x\rangle_P|B'\rangle_P)
\end{equation}

\vspace{5mm}

Bell explained that if one measures the properties of $P$ in region $I$,
then one would observe interference phenomena (in this experiment, for
example, one would observe $z$-spin up with probability one).  The
observation of interference phenomena is usually taken to entail
that $P$ could not have followed path $A$ and could not have followed
path $B$ since, in either case, the probably of observing $z$-spin up
in $I$ would presumably be $1/2$ (as predicted by the standard
collapse formulation of quantum mechanics).  In
such a situation, one would say, on the standard view,
that $P$ followed a {\em superposition\/} of
the two trajectories (which, on the standard interpretation of
states is supposed to be neither one nor the other
nor both trajectories).  But according to Bohm's theory, $P$ determinately
follows one or the other of the two trajectories: that is, it either
determinately follows $A$ or it determinately follows $B$.  On Bohm's theory,
one might say that the interference
effects that one observes at $I$ are the result of the wave function
following both paths.

If we do not observe the particle in region $I$, then $P$ will arrive
at one of the two detectors to the right of
the interference region: either the one at $A'$ or the one at $B'$.  If
it arrives at $A'$, then one might suppose that the particle traveled
path $A$; and if it arrives at $B'$, then one might suppose that it
traveled path $B$.  But such inferences do not work in the standard collapse
formulation of quantum mechanics, where (according to the standard
eigenvalue-eigenstate link) under these circumstances $P$ would
have traveled a superpostion of the two paths.  And such inferences do not work in Bohm's theory either,
but for a very different reason.  In Bohm's theory the particle really does
travel one or the other of the two paths, it is just that its trajectory is
not what one might at first expect.

In figuring out what trajectory Bohm's theory predicts, the
first thing to note is that, by symmetry, the probability current
across the line $L$ is always zero.\footnote{See Phillipidas, Dewdney, and
Hiley (1979) for an explicit calculation of the trajectories in for a similar
experiment.  The explicit calulations, of course, show that possible particle
trajectories never cross $L$.  Bell cites this paper at the end of his 1980
paper on delayed-choice experiments in Bohm's theory.}
This means that if $P$ starts in the top half of the initial wave packet, then it
must move from $S$ to $A$ to $I$ to $B'$; and if it starts in the bottom half of the
initial wave packet, then it must move from $S$ to $B$ to
$I$ to $A'$.  That is, whichever path $P$ takes, Bohm's theory
predicts that it will {\em bounce\/} when it gets to region $I$---in order
to follow either trajectory, the particle $P$ must accelerate in the 
field-free region $I$.

Concerning this odd bouncing behavior Bell says that ``it is vital here
to put away the classical prejudice that a particle moves in a 
straight path in `field free' space'' (1987, 113).  But certainly, one might
object, this is more than a prejudice.  After all, this particle bouncing
nonsense is a direct violation of the conservation of momentum, and we have very
good empirical reasons for supposing that momentum is conserved.  Isn't
this alone reason enough to dismiss Bohm's theory?  Put another way,  whenever we 
observe which path the particle
in fact travels, if we find it at $A'$, then we also observed it traveling path
$A$ and if we find it at $B'$, then we also observed it traveling path $B$.
That is, whenever we make the appropriate observations, we never observe the
crazy bouncing behavior (or any of the other violations of the conservation
of momentum) predicted by Bohmian mechanics.

This puzzling situation is the basis
for ESSW's surreal trajectories argument.  The
argument goes something like this:

\vspace{5mm}

\noindent
{\bf Assumption 1} (explicit):  Our experimental measurement records tell us that in a two-path
interference experiment like that described above each particle either travels from
$A$ to $A'$ or from $B$ to $B'$; that is, they never bounce.
\vspace{.5cm}

\noindent
{\bf Assumption 2} (implicit):  Our measurement records reliably tell us where a particle is at the
moment the record is made.
\vspace{.5cm}

\noindent
{\bf Assumption 3} (implicit):  One can record which path a particular particle takes without
breaking the symmetry in the probability currents that prevents the particle from crossing the line
$L$.

\vspace{5mm}

\noindent
{\bf Conclusion}: The trajectory predicted by Bohm's theory, where the particle bounces, cannot
be the particle's actual trajectory; that is, Bohm trajectories are not real, they are ``surreal.''   And
if the trajectories predicted by Bohm's theory are not the actual particle trajectories, then Bohm's
theory is false, and this constitutes very good grounds for rejecting it.

\vspace{5mm}

D\"urr, Fusseder, Goldstein, and Zanghi (DFGZ) immediately responded 
to defend Bohm's theory against the surreal trajectories argument:
\begin{quote}
In a recent paper [ESSW (1992)] it is argued that dispite its many
virtues---its clarity and simplicity, both conceptual and physical, and the
fact that it resolves the notorious conceptual difficulties which plague
orthodox quantum theory---BM [Bohmian mechanics] itself suffers from a fatal
flaw: the trajectories that it defines are ``surrealistic''.  It must be
admitted that this is an intriguing claim, though an open minded advocate
of quantum orthodoxy would presumably have preferred the clearer and
stronger claim that BM is {\em incompatible\/} with the predictions of
quantum theory, so that, despite its virtues, it would not in fact provide
an explanation of quantum phenomena.  The authors are, however, aware
that such a claim would be false. (1993, 1261)
\end{quote}
And since Bohm's theory makes the same predictions as the standard
theory of quantum mechanics, DFGZ argue that ESSW
cannot possibly provide, as ESSW describe it, ``an experimentum crucis which, according
to our quantum theoretic prediction, will clearly demonstrate that the
reality attributed to Bohm trajectories is rather metaphysical than
physical.''  And with this DFGZ dismiss ESSW's
argument against Bohmian mechanics:
\begin{quote}
On the principle that the suggestions of scientists who propose pointless
experiments cannot be relied upon with absolute confidence, with this proposal
the [ESSW] paper self-destructs: The authors readily agree that
the ``quantum theoretical predictions'' are also the predictions of BM.  Thus
they should recognize that the [experimental] outcome on the basis
of which they hope to discredit BM is precisely the outcome predicted by
BM.  Under the circumstances it would appear prudent for the funding agencies
to save their money! (1261)
\end{quote}
DFGZ conclude their defense of Bohm's theory by making a point about the
theory-ladenness of talk of particle trajectories and a point about
the theory-ladenness of observation itself.  But we will return to these
two (important) points later, when we have the conceptual tools
hand to make sense of them (Section~5).

In their reply to DFGZ's comment, ESSW want to make it perfectly clear that
they did not anywhere conceed that Bohm's theory had ``many
virtues'' nor did they admit that the orthodox formulation of quantum mechanics
was ``plagued by notorious conceptual difficulties.''  But, for their part, ESSW do
seem to conceed, as DFGZ insisted, that Bohmiam mechanics makes the same empirical predictions
as standard quantum mechanics: ``Nowhere did we claim that BM makes
predictions that differ from those of standard quantum mechanics''
(1263).\footnote{But ESSW later make the following argument in favor
of actually funding the surreal trajectories
experiments that they describe: ``Funding agencies were and are well advised to support experiments that
have probed or would probe the ``surprises'' of quantum theory.  Imagine
the (farfetched) situation that the experimenter finds the photon always
in the resonator through which the Bohm trajectory passes rather than the
one predicted by quantum theory.  Wouldn't that please the advocates of
BM?'' (1263--4).  This is, of course, very puzzling talk indeed once
EWWS conceed that Bohmian mechanics makes the same empirical predictions as
the standard theory---a proponent of Bohm's theory would
most certainly {\em not\/} be pleased if experiments showed that the standard
quantum-mechanical predictions were false because this would mean
that Bohm's theory was itself false!  When ESSW say things like this, it
is easy to understand DFGZ's frustration.}  Rather than argue that
Bohm's theory made the wrong empirical predictions, ESSW claim that
the purpose of their original paper was ``to show clearly that the
interpretation of the Bohm trajectory---as the real retrodicted history
of the [test particle that travels through the interferometer]---is
implausible, because this trajectory can be macroscopically at
variance with the detected, actual way though the interferometer''
(1263).  This last clause identifies the detected path with
the actual path traveled by the test particle.  This is their (implicit)
assumption that particle detectors would be reliable (in a perfectly
straightforward way) on the delayed-choice interference experiments
that they discuss.  ESSW conclude, ``Irrespective of what can be said in
addition, we think that we have done a useful job in demonstrating
just how artificial the Bohm trajectories can be'' (1264).

Again, ESSW's claim is not that Bohm's theory makes the wrong empirical
predictions nor it is that the theory is somehow logically inconsistent; rather, they
argue (on the implicit assumption that our particle detectors reliably tell us where
particles are) that Bohm's theory makes the wrong predictions for the actual motions
of particles---that the predicted particle trajectories are ``artificial,''
``metaphysical,'' and, at best, ``implausible.''

While I agree with DFGZ that surreal trajectories are not something that a proponent
of Bohm's theory should worry about, the full story is a bit more involved
than the sketch given in their comment on ESSW's paper.  In order
to get everything straight, let's return to Bell's original analysis
of the delayed-choice interference experiment in the context of Bohm's
theory.

\section{}

Bell's analysis of the delayed-choice interference
experiment provides a good first step in explaining why
conservation-of-momentum-violating ``surreal'' trajectories
do not pose a problem for Bohm's theory.
While Bohm's theory does indeed predict that momentum (in the usual
sense) is not conserved in experiments like that described above,
Bell explained why one would never detect violations of the
conservation of momentum.  The short story is this: while the {\em 
actual\/} momentum (mass times particle velocity) is typically not
conserved in Bohm's theory, the {\em measured\/}
momentum (as expressed by the results of what one would ordinarily
take to be momentum measurements) is always conserved.

In order to detect a momentum-violating bounce in an experiment
like that described above, one would have to perform {\em two\/}
measurements: one to show which path the particle travels,
($A$ or $B$) and another to show where the particle ends up ($A'$ or $B'$).  One
might then try to show that a particle that travels path $A$, say, ends up at $B'$,
and thus violates the conservation of momentum.  But one will never
observe such a bounce in Bohm's theory {\em because measuring
which path the particle follows will destroy the symmetry in the probability
currents that generate the bounce}.  That is, particles only exhibit their
crazy bouncing behavior in Bohm's theory when no one is looking!

Suppose (following Bell 1980) that one puts a detector on path
$B$ designed to correlate the position of a flag with the position of the
test particle $P$ (see figure~2).  More
specifically, consider a single flag particle $F$ whose position
(as represented by the quantum-mechanical state)
gets correlated with the position of $P$ as follows: (1) if $P$
is in an eigenstate of traveling path $A$, then $F$ remains in an eigenstate
of pointing at ``No'' and (2) if $P$ is in an eigenstate of traveling
path $B$, then $F$ ends up in an eigenstate of pointing at
``Yes''.  That is, the detector is designed so that the position of $F$
will record the path taken by $P$.

\vspace{5mm}
[Figure 2: Experiment where $F$'s position is correlated with the position of $P$]
\vspace{5mm}

While this experiment may look like the earlier one, introducing
such a detector requires one to tell a very different story
than the one told without the detector. 

Given the nature of the interaction between $P$ and $F$ and the
linearity of the dynamics, if $P$ begins in the $z$-spin up state
(a superposition of $x$-spin eigenstates), then the effective wave
function of the composite system would evolve as follows:

\vspace{5mm}

\noindent
Initial state:
\begin{equation}
|\upa_z\rangle_P|S\rangle_P |\mbox{``No''}\rangle_F =
|S\rangle_P|\mbox{``No''}\rangle_F 1/\sqrt{2}
(|\upa_x\rangle_P + |\dna_x\rangle_P)
\end{equation}

\noindent
$P$'s wave packet splits:
\begin{equation}
|\mbox{``No''}\rangle_F 1/\sqrt{2} (|\upa_x\rangle_P|A\rangle_P +
|\dna_x\rangle_P|B\rangle_P)
\end{equation}

\noindent
$M$'s position is correlated with the position of $P$:

\begin{equation}
1/\sqrt{2} (|\upa_x\rangle_P|A\rangle_P
|\mbox{``No''}\rangle_F + |\dna_x\rangle_P|B\rangle_P|\mbox{``Yes''}\rangle_F)
\end{equation}

\noindent
The two wave packets appear to pass though each other in region $I$ (but they {\em miss\/} each other in
configuration space!):
\begin{equation}
1/\sqrt{2} (|\upa_x\rangle_P|I\rangle_P |\mbox{``No''}\rangle_F +
|\dna_x\rangle_P|I\rangle_P|\mbox{``Yes''}\rangle_F)
\end{equation}

\noindent
Final state:
\begin{equation}
1/\sqrt{2} (|\upa_x\rangle_P|A'\rangle_P |\mbox{``No''}\rangle_F +
|\dna_x\rangle_P|B'\rangle_P|\mbox{``Yes''}\rangle_F)
\end{equation}

\vspace{5mm}

Note that the position of $F$ does in fact reliably record where $P$ was when
the position record was made.  Because the wave function associated with the
two possible positions for $F$ do not overlap in configuration
space, the position correlation between $P$ and $F$ destroys the symmetry that prevents
$P$ from crossing $L$.  While the two wave packets both appear to pass through region $I$
at the same time, they in fact
miss each other in configuration space.  In order to see how $P$ and $F$ move,
consider the evolution of the wave function and the two-particle configuration in
configuration space.

\vspace{5mm}
[Figure 3: The Last Experiment in Configuration Space]
\vspace{5mm}

If the two-particle configuration starts in the top half of the initial wave packet
(as represented in Figure 3), then $P$ would
move from $S$ to $A$ to $I$ to $A'$ and $F$ would stay at ``No''.  If the configuration starts in
the bottom half of the initial wave packet, then $P$ would move from $S$ to $B$ then $F$
would move to ``Yes'' then $P$ would move from $B$ to $I$ to $B'$.  That is, regardless of
where $P$ starts, it will pass though the region $I$ without bouncing.  Moreover, $F$ will
record that $P$ was on path $A$ if and only if $P$ ends up at $A'$ and that $P$ was on path
$B$ if and only if $P$ ends up at $B'$.  That is, if one makes a determinate record of $P$'s
position before $P$ gets to $I$, then $P$ will follow a perfectly natural trajectory, and the record
will be reliable.  Again, recording the position of $P$ destroys the symmetry that
prevents $P$ from crossing $L$.

This experiment illustrates why a measurement record is reliable in Bohm's theory whenever
there is a strong correlation between the position of the system being observed and the
position of the recording system.  And since all measurements are ultimately position
measurements on the minimal Bohm's theory, one might
simply conclude that all determinate records produced by strong correlations are reliable in
Bohm's theory and dismiss the surreal-trajectories problem as a problem
that was solved by Bell before it was even posed by ESSW.  This is not such a bad
conclusion, but the right thing to say about surreal trajectories is
slightly more subtle.

Note that in order to tell a story like the one above, one must record
the path taken by the test particle in terms of the position of
something.  Here the record is in terms of the position of the
flag particle $F$.  It is this position correlation that breaks the symmetry
in the probability currents, which then allows the test particle
$P$ to follow a momentum-conserving
trajectory.  All it takes is a strong position correlation with even
a single particle.  And it is this that makes the final position record
reliable.\footnote{Bell explained that a good measurement
record must make a macroscopic difference.  He emphasized that a discharged
detector is macroscopically different from an undischarged detector.  This is also
something emphasized by DFGZ (1993, 1262) in order to argue that one would
not expect one of ESSW's detectors to generate a sensible record of which path the
test particle followed "until an interaction with a suitable macroscopic device
occurs."  But note that all that really matters here is that the wave packets that
correspond to different measurement outcomes (in terms of the position
of $F$) be well-separated in configuration space in the $F$-position-record
direction.  This is not to say that Bell's (and DFGZ's) point concerning macroscopic
differences irrelevant.  If the flag is a macroscopic system that makes a macroscopic movement,
then this will obviously help to provide the wave packet separation required for a
reliable record.  But while a macroscopic position correlation with a macroscopic
system sufficient, it is not a necessary condition for generating a reliable record
in Bohm's theory.  See Aharonov, Y.\ and L.\ Vaidman (1996) for a discussion of
partial measurements in Bohm's theory, measurements where the separation
between the post-measurement wave packets in configuration space is
incomplete.}

So what would happen if one tried to record the position of $P$ in terms
of some physical property other than position?  This is something that is
important to our making sense of the history of the surreal trajectories
problem, but it is something that Bell did not consider.

\section{}

In order to avoid Bell's (preemptive) dissolution of the surreal trajectories
problem  ESSW must have had in mind a different sort of which-path
detector than the one considered by Bell. Indeed, the
experiments that ESSW describe in their 1992 paper employ detectors
that record the path followed by the test particle in the
creation of photons.  A proponent of Bohm's theory might
point out that since the theory is explicitly
nonrelativistic and since the very statement of its auxillary dynamics
requires there to be a fixed number of particles, these experiments
are simply outside the domain of the theory.
But perhaps it is possible to capture at least the spirit of ESSW's experiments with
experiments that are well within the domain of the minimal Bohm's theory.\footnote{The experiment
below shows what would happen if one tried to record the path taken by the particle in
the $x$-spin of another particle.  Dewdney, Hardy, and Squires (1993) tried to capture
the spirit of ESSW's experiments by showing in graphic detail what would
happen if one tried to record the path in terms of energy.}

Consider what happens when one {\em tries\/} to record $P$'s position in something other than
position (one might naturally, and quite correctly, object that there is no other
quantity in Bohm's theory that one {\em could\/} use to record $P$'s position, but with
the aim of trying to revive the surreal trajectories problem, read
on).  Suppose, for example, that one tries to record $P$'s position in a particle $M$'s $x$-spin:
that is, suppose that the interaction between $P$ and $M$ is such that if $P$'s initial
effective wave function were an $x$-spin up eigenstate, then nothing would happen to $M$'s
effective wave function; but if $P$'s initial effective wave function were an $x$-spin down
eigenstate, then the spin index of $M$'s effective wave function would be flipped from $x$-spin
up to $x$-spin down (since $x$-spin is a contextual property in the minimal Bohm's theory, the value
of the $x$-spin record depends, as we will see, on how it is read, and one
might thus, quite correctly, argue that it is not a record of the position
of $P$ at all, but read on).  In the standard von Neumann-Dirac collapse formulation of quantum
mechanics (once a collapse had eliminated one term or the other of the correlated superposition!)
one might naturally think of this interaction as recording $P$'s position in $M$'s $x$-spin.  On
this view, $M$ might be thought of as a sort of which-path detector.

Continuing with the experimental set up,
suppose further that $M$'s $x$-spin might then be converted to a position
record by a detector with a flag particle $F$ designed to
point at ``No'' if $M$ is in the $x$-spin up state and to point at
``Yes'' if $M$ is in the $x$-spin down state.  The conversion of the
$x$-spin record (though it will turn out that there is no determinate
$M$-record {\em until after this conversion is made\/} in the delay-choice
interference experiments!) to a
position record here consists in correlating
the position of $F$ with the $x$-spin of $M$ (as represented by the
quantum-mechanical state).  The idea is that if $P$ is
in an eigenstate of traveling path $B$, say, then $M$ will record
this fact by the $x$-spin index on its wave function being
flipped to $x$-spin down, which is something that might then
be converted into a record in terms of the position of $F$ through
the interaction between $M$ and $F$.  In this case, $F$ would move to record
the measurement result ``Yes''.

\vspace{5mm}
[Figure 4: One tries to record the position of $P$ in the $x$-spin of $M$]
\vspace{5mm}

The effective wave function of the composite system then evolves as follows:

\vspace{5mm}

\noindent
Initial state:
\begin{equation}
|\upa_z\rangle_P|S\rangle_P |\upa_x\rangle_M |\mbox{``No''}\rangle_F =
|S\rangle_P |\upa_x\rangle_M|\mbox{``No''}\rangle_F 1/\sqrt{2} (|\upa_x\rangle_P +
|\dna_x\rangle_P)
\end{equation}

\noindent
$P$ wave packet is split:
\begin{equation}
|\upa_x\rangle_M|\mbox{``No''}\rangle_F 1/\sqrt{2}
(|\upa_x\rangle_P|A\rangle_P + |\dna_x\rangle_P|B\rangle_P)
\end{equation}

\noindent
The $x$-spin component of $M$'s wave packet is correlated to the position of
$P$'s:
\begin{equation}
|\mbox{``No''}\rangle_F 1/\sqrt{2} (|\upa_x\rangle_P|A\rangle_P|\upa_x\rangle_M +
|\dna_x\rangle_P|B\rangle_P|\dna_x\rangle_M)
\end{equation}

\noindent
The two wave packets pass through each other in configuration space:
\begin{equation}
|\mbox{``No''}\rangle_F
1/\sqrt{2} (|\upa_x\rangle_P|I\rangle_P|\upa_x\rangle_M +
|\dna_x\rangle_P|I\rangle_P|\dna_x\rangle_M)
\end{equation}

\noindent
Then they separate:
\begin{equation} 
|\mbox{``No''}\rangle_F 1/\sqrt{2}
(|\upa_x\rangle_P|A'\rangle_P|\upa_x\rangle_M +
|\dna_x\rangle_P|B'\rangle_P|\dna_x\rangle_M)
\end{equation}

\noindent
Then the position of the $F$ is correlated to
the $x$-spin component of $M$'s wave packet:
\begin{equation}
1/\sqrt{2}
(|\upa_x\rangle_P|A'\rangle_P|\upa_x\rangle_M |\mbox{``No''}\rangle_F +
|\dna_x\rangle_P|B'\rangle_P|\dna_x\rangle_M
|\mbox{``Yes''}\rangle_F)
\end{equation}

\vspace{5mm}

Note that here the symmetry in the probability current that prevents $P$ from crossing $L$ is
preserved.  That is, $P$ bounces just as it did in the first experiment
we considered.

\vspace{5mm}
[Figure 5: Last experiment in configuration space]
\vspace{5mm}

If the three-particle configuration begins in the top half of the initial wave packet (as
represented in Figure~5), then $P$ will move from $S$
to $A$ to $I$ to $B'$ then, when $M$ and $F$ interact, $F$ will move to ``Yes''.  If the three-particle configuration begins in
the bottom half of the initial wave packet, then $P$ will move from $S$ to $B$ to $I$ to $A'$ and, when
$M$ and $F$ interact, $F$ will stay at ``No''.  That is, the final position of the $F$ will be at ``No'' if $P$ traveled
along the lower path and it will be at ``Yes'' if $P$ traveled along the upper path.  In other words, $F$'s
final position does not tell us which path $P$ followed in the way that it was intended.

One might naturally conclude that the which-path detector is fooled by the late measurement, and
defend Bohm's theory against ESSW by denying their implicit assumption that
the which-path detectors are reliable.\footnote{In their 1993 paper, Dewdney, Hardy, and Squires
argue precisely this.}  There is, however, another way of looking at a delayed-choice
interference experiment where one tries to record the path in some property other
than position.  One might claim both that Bohm's theory is true {\em and\/}
that one's detectors are perfectly reliable.  This, it seems to me, is an option
suggested by DFGZ's discussion of the theory-ladenness of talk of trajectories
and of observation itself near the end of their response to ESSW's original
paper.

Given their contention that the experiments described by ESSW
could provide no empirical reason for rejecting Bohmian mechanics, DFGZ
ask the question ``So what on earth is going on here?''
\begin{quote}
The answer appears to be this: The authors [ESSW] distinguish between the Bohm
trajectory for the atom and the {\em detected\/} path of the atom.  In
this regard it would be well to bear in mind that before one can speak
coherently about the path of a particle, detected or otherwise, one
must have in mind a theoretical framework in terms of which this notion
has some meaning.  BM provides one such framework, but it should be clear
that within this framework the [test particle] can be detected passing
only through the slit through which its trajectory in fact passes.  More
to the point, within a Bohmian framework it is the very existence of
trajectories wich allows us to assign some meaning to this talk about
detection of paths. (1261--2)
\end{quote}
It seems that there are two points here.  The first
point concerns the theory-ladenness of talk about trajectories.  On
the orthodox formulation of quantum mechanics,
there is no matter of fact at all concerning which path the
test particle traveled since it simply fails to have any
determinate position whatsoever before it is detected.  Indeed,
insofar as ESSW's description of the surreal trajectories experments
presupposes that there are determinate particle
trajectories, they are presupposing something that is {\em incompatible\/}
with the very quantum orthodoxy they seek to defend!  The point
here is that any talk of
determinate trajectories is talk within a theory.  A precondition
of such talk is that one have a theory where there are
determinate trajectories,
a theory like Bohmian mechanics.

DFGZ's second point, if I understand it correctly, concerns
the theory-ladenness of observation,
but this will first require some clarification.  Their claim that
in Bohmian mechanics a test particle ``can
be detected passing only through the slit through which its trajectory
in fact passes'' suggests that they were considering only experiments
like those in the last section where the which-path detector indicates
in a perfectly straighforward way the path that the test particle
in fact followed.  But DFGZ do in fact grant that there
are situations where Bohm's theory predicts that a
late observation of a which-path detector would find that
the detector registers that the test particle traveled one
path when it in fact traveled the other.  They also grant
that this is somewhat surprising.  But
they explain that ``if we have learned anything by now about
quantum theory, we should have learned to expect surprises!''  And
DFGZ maintain that even in such experiments the measurement performed
by the which-path detector ``can indeed be regarded as a measurement of
which path the [test particle] has taken, but one that conveys information
which contradicts what naively would have been expected.''  DFGZ then draw
the moral that ``BM, together with the authors [ESSW] of the paper
on which we are commenting, does us the service of making it dramatically
clear how very dependent upon theory is any talk of measurement or
observation'' (1262).

While it is not entirely clear what DFGZ have in mind, one
way to read this is that,
contrary to what is later argued by Dewdney, Hardy, and
Squires (1993), DFGZ take the which-path detector to be perfectly
reliable even in experiments where it ``records'' that the test particle
traveled one path when it in fact (according to Bohm's theory) traveled
the other {\em once one understands what the detector is detecting}.
On this reading, then, the point here is that since observation is
itself a theory-laden notion, what one is detecting can only be determined
in the context of a theory that explains what it is that one is detecting.
But if this is what DFGZ had in mind, then
what exactly does Bohm's theory tell us that a which-path detector is
detecting in the context of a late-measurement experiment? (And is it really
possible to tell a plausible story where the detectors are perfectly
reliable here?)

Perhaps the easiest answer would be to insist that when one tries
to record the path taken by the test particle in a property other than position (in a
delayed-choice interference experiment), one's which-path detector simply works
in exactly the opposite way that one would expect.  The detector is
perfectly reliable---it is just that when it records that the test particle
traveled path $A$, the detector record (under such circumstances) really {\em means},
according to Bohm's theory, that the test particle in fact traveled path $B$;
and, similarly, on this view a $B$ record {\em means}, according to Bohm's theory,
that the test particle traveled path $A$.

It seems, however, that this cannot be quite right.
When one tries to record the path that the test particle 
traveled in a property other than position (in the
delayed-choice interference
experiment), there is no determinate record whatsoever (on the minimal Bohm's theory)
before the test particle passes through the interference region $I$ because
the which-path detector has not yet correlated the position of anything
with the position of the test particle.

The right thing to say, it seems to me,
is that while 
the which-path detector does not detect anything before one
correlates the position of the flag $F$ with $M$'s $x$-spin (on
the minimal Bohm's theory), whenever one makes a
determinate record in Bohm's theory using a device that
induces a strong correlation between the measured position of
the object system and the position that records the outcome, then
that record will be perfectly reliable {\em at the moment the
determinate record is made}.  On this view, there is still a sense
in which one can think of the
detectors in the delayed-choice interference experiments as
being perfectly reliable, but this will take some explaining.

As DFGZ suggest, we naturally rely on our best physical theories to tell us
what it is that our measuring
devices in fact measure, so what does Bohm's theory tell us about the
{\em late-measurement\/} of the which-path detector in the delayed-choice
interference experiment?  Note, again, that there is no determinate
record whatsoever before the late measurement.  Also note that
while the final position of $F$ does not
tell us where $P$ {\em was\/} when $P$ interacted with $M$,
it does reliably tell us where $P$ {\em is\/} at the moment that
the x-spin correlation is converted into a determinate measurement
record (when the position
of $F$ is correlated with the $x$-spin of $M$): if one gets the
result ``No'' ($x$-spin up), then the theory tells us that $P$ is currently
associated with the $x$-spin up wave packet {\em wherever that wave 
packet may be}, and if one gets the result ``Yes'' ($x$-spin down), then it tells
us that $P$ is currently associated with the $x$-spin down wave packet {\em wherever that wave 
packet may be}.

So this is how it works.  Since the
only determinate noncontextual records in Bohm's theory are records
in terms of the position of something,
there is, stictly speaking, no determinate record of $P$'s position until we convert
the correlation between the position of $P$ and the $x$-spin of $M$ into a correlation between the
position of $P$ and the position of $F$.  And whenever this position correlation is made,
we reliably, and {\em nonlocally}, generate a record of $P$'s position {\em at that moment\/}.
If we wait until after $P$ has passed through region $I$, then
if $F$ stays at ``No'', this means that $P$ is associated with the $x$-spin up component
which means that it is at position $A'$, and if $F$ moves to ``Yes'', this means that $P$ is
associated with the $x$-spin down component which means that it is at position $B'$.
The moral is that one cannot use a record in Bohm's theory to figure out which path
$P$ took unless one knows how and when the record was made.

But note that in this Bohm's
theory is arguably better off than the standard von Neumann-Dirac
collapse formulation of quantum mechanics.  On the standard
eigenvalue-eigenstate link (where a system determinately has a property if and only if
it is in an eigehstate of having the property)
one can say {\em nothing whatsoever\/} about which trajectory
a particle followed since it would typically fail to have {\em any\/}
determinate position until it was observed.  If one does not worry
about the unreliability of retrodiction in the
context of the standard collapse theory (and ESSW do not seem 
to be worried about this!), then I can see no reason at all to worry about
it in the context of Bohm's theory.  Further, there is no reason to suppose that Bohmian particle trajectories are not the actual
particle trajectories.  Nor is there any reason to conclude that our good particle
detectors are somehow unreliable.  Rather than saying that a detector is fooled by
a late measurement, one should, I suggest, say that the late measurement {\em reliably\/}
detects the position of the test particle {\em nonlocally}. 

On this view the surreal trajectories experiments simply serve to reveal the
special role played by position and, ultimately, the nonlocal
structure of Bohm's theory.  As Bell explained, ``The
fact that the guiding wave, in the general case, propagates not
in ordinary three space, but in a multi-dimentional configuration
space in the origin of the notorious `nonlocality' of quantum mechanics.
It is a merit of the de Broglie-Bohm version to bring this out so
explicity that it cannot be ignored'' (1987, 115).  But one should
note that it is not some subtle sort of nonlocality involved
in the account of quantum-mechanical correlations here.  The
configuration space particle dynamics that
accounts for the nonlocal correlations in the late-measurement
experiments makes Bohm's theory incompatible
with relativity.

But there is one more point that I would like to
make before turning to a discussion of the relationship between
how Bohm's theory accounts for surreal trajectories and its
incompatibility with relativity.  As suggested above (on
the minimal Bohm's theory) whenever the position
of one system is recorded in the position of another system via a strong
correlation between the effective wave functions of the two systems (one
that produces the appropriate separation of the wave function in the
recording parameter in configuration space), then that record will
reliably indicate where the measured particle is {\em at the moment
the determinate record is made}.  It is also the case (on the
minimal Bohm theory) that all determinate records are ultimately
position records.  One can only take these facts to
provide a solution to the surreal trajectories if one allows for
Bohm's theory to tell one something about what one is observing
when one observes (or, in somewhat different language, what
constitues a good measuring device).  But
it seems that this is precisely the sort of thing that one must be
willing to do when entertaining a new theoretical option.
One might dogmatically insist on holding to one's pre-theoretic
intuitions concerning what one's detectors detect come what may,
but this would certainly be a methodological mistake.

\section{}

Consider again the late-measurement experiment of the last section (see
figures~4 and~5).  If $P$ begins in the top
half of the wave function at $S$, it will travel path $A$ to $I$ in the
$x$-spin up wave packet.  That is, before $P$ gets to $I$, the
three-particle configuration will be associated with the $x$-spin up
component of the wave function in configuration space.  And this means
that if one converts the spin record into a position record {\em before\/}
the two wave packets interfer at $I$, one will get the result ``No''.  But
if $P$ continues to $I$, bounces, and the two-particle configuration
is picked up by the $x$-spin down wave packet, then, since the two-particle
configuration is now associated with the $x$-spin down wave packet, if
one now converts the $x$-spin record into a position record, one
will get the result ``Yes'' .

This means that one might instantaneously
determine the value of the converted record at $B$ (the record
one gets by converting the $M$ $x$-spin ``record'' into an $F$
position record) by choosing whether or not to interfer the two wave
packets at $I$.  If the two wave packets pass through each other,
then $F$ will move to ``Yes'' when the spin record is
converted; if not, then $F$ will stay at ``No'' when the spin record is converted.  So,
if someone at $I$ knew which path $P$ was on (something, as explained
earlier, that is prohibited in Bohm's theory if the distribution postulate is
satisfied), then he or she could use this information to send a superluminal
signal to a friend on path $B$ by deciding whether or
not to interfere the wave packets at $I$.  But regardless
of whether one knows which
path $P$ is on, the theory predicts (insofar as one is comfortable with
the relavant counterfactuals in the context of a deterministic theory)
that one can instantaneously affect the result of a
measurement of $M$ from region $I$, and one might take the possibility
of superluminal effects here to illustrate the
incompatibility of Bohm's theory and relativity.

This incompatibility is more clearly illustrated by considering the
role that the
temporal order of events plays in Bohm's theory.  Consider
the late-measurement experiment one more time.  
If one converts the spin record 
{\em before\/} the two wave packets interfer at $I$, then one will get
the result ``No''; and if one converts the spin record {\em after} the
wave packets interfer, then one will get the result ``Yes''.  But if the
conversion of the spin record and the interference
of the wave packets are space-like separated events, then
the conversion event occurs before the interference
event in some inertial frames and after the interference event in others.
So in order to get any empirical predictions whatsoever out of Bohm's
theory for this experiment whenever the conversion 
and interference events are space-like separated, one must choose
a perfered inertial frame that imposes a perfered temporal order
on the conversion and interference events.  But having to choose a perfered
intertial frame here is a direct violation of the basic principles
of relativity. This is the sense in which the account that Bohm's theory
provides of the late-measurement experiment 
is fundamentally incompatible with relativity.

If the distribution postulate is satisfied, then Bohm's theory
makes the same empirical predictions as the standard von Neumann-Dirac formulation
of quantum mechanics (whenever the latter makes unambiguous predictions)
and the standard quantum statistics do not allow one to 
send superluminal messages (given the usual quantum statistics,
one can prove a no-signaling theorem).  So while Bohm's
theory is not Lorentz-covariant, it explains why one would
never notice this fact (just as it explains why one would never
notice violations in the conservation
of momentum).

A proponent of Bohm's theory might argue that nonlocally
correlated motions
like the correlated motions in the conversion and
interference events describe above is too weak of a relationship
to be causal, and
that Bohm's theory thus does not in fact allow
for nonlocal causation.  While
such a conclusion would do nothing to make Bohm's theory compatible with relativity even
if it were granted, I do not think that it should be granted.  It seems to me that
if any correlated motions should count as causally connected in Bohm's theory,
then nonlocal correlated
motions should as well.  Nonlocal correlated motions, like local correlated
motions (insofar as their are any truly {\em local\/} correlated motions!), are simply
the result of the configuration
space evolution of the physical state.  The point here is that Bohm's theory handles nonlocal
correlated motions precisely the same way that it handles events that one would
presumably want to count as causal---like the correlated motion produced
between a football and the foot that kicks it. Of course, one might resist the conclusion
that nonlocal correlated motions are causally related by denying that
there are any causal relationships whatsoever in Bohm's theory.  But this
would mean that even those explanations that one gives 
that look like causal explanations are not, and this
seems to me to be putting things the wrong way around.  Just
as we look to our best theories to tell us how to build
good detectors and to explain what it is that they
detect, it seems that we should also look to
our best theories to tell us something about the nature of
causal relations.  There is nothing
inherently wrong with sitting down and deciding once and for all
the necessary and sufficient conditions for events to be
causally related.  It is just that one risks adopting a notion
of causation that is irrelavant to the sort of explanations provided
by our best physical theories.\footnote{Michael Dickson (1996) has argued
that it does not make any sense to ask whether a deterministic
theory like Bohm's theory is local because of the difficulty supporting
counterfactual conditionals in such a theory.  He also suggests that
the notion of causality may not make make sense in such
a theory either (1996, 329).  I agree
that some intuitions concerning what it would mean for a theory
to be local or what it would mean for one event to cause another
cannot be supported in a deterministic theory, but this
does not mean that we can make no sense at all of what it would
be for a deterministic theory to be local or for one event to cause
another.  Indeed, whether Bohm's theory is Lorentz covariant is a
perfectly good sensible question concerning its locality---and
it isn't.}

But regardless of what one thinks about causality,
the particle trajectories predicted by Bohm's theory depend
on one's choice of
inertial frame, which means that the theory is incompatible with
the basic principles of relativity.  And this is the real
problem.

\section{}

It is the configuration space dynamics that makes
Bohm's theory incompatible with relativity.  But it
is also the instantaneous correlated
motion predicted by the configuration space dynamics
that explains the quantum-mechanical correlations in Bohm's
theory and makes the theory empirically adequate.  And it is
the configuration space dynamics that allows one to say that
whenever the position of one system is
recorded in the position of another system via a strong correlation between the
effective wave functions of the two systems, then that
record will reliably indicate where the
measured system is {\em at the moment the determinate
record is made}, which, it seems to me, is ultimately the best
response to the supposed surreal trajectory problem.

But this leaves a proponent of Bohm's theory with a difficult
choice.  One might try to find some new way to account for
quantum-mechanical correlations, one that does not require
a preferred temporal order for space-like separated events where
objects exhibit correlated properties.  But it should 
be clear from the the configuration-space stories
told above that such a theory would have to explain
quantum-mechanical correlations in a way that is fundamentally 
different from the configuration-space way in which they are
explained by Bohm's theory.  And, of course, actually finding
such an alternative is much easier said
than done.\footnote{For other discussions of the incompatibility
of Bohm's theory and relativity see Albert (1992) and Artnzenius (1994).
For a recent discussion concerning the difficulty in getting a
Bohm-like auxiliary quantum dynamics that is compatible with relativity
see Dickson and Clifton (1998).}  Or one might simply drop
the requirement of Lorentz covariance as a feature of
a satisfactory dynamics and settle for something weaker, perhaps
something like {\em appearant\/} Lorentz covariance.  But this would
be an enormous theoretical sacrifice---presumably one that few physicists would
seriously entertain.

\newpage
\begin{center}
\Large
REFERENCES
\normalsize
\end{center}
\vspace{.5in}

\vspace{5mm}
\noindent
Aharonov, Y.\ and L.\ Vaidman: 1996, ``About Position
Measurements which do not show the Bohmian Particle Position,''
in J.\ T.\ Cushing et al.\ (eds), (1996, 141--154).

\vspace{5mm}
\noindent
Albert, D.\ Z.: 1992, {\em Quantum Mechanics and Experience},
Harvard University Press, Cambridge.

\vspace{5mm}
\noindent
Arntzenius, F.: 1994, `Relativistic Hidden-Variable
Theories?' {\em Erkenntnis\/}
41, 207--231.

\vspace{5mm}
\noindent
Bacciagaluppi, G.\ and M.\ Dickson: 1996, `Modal Interpretations
with Dynamics' In Dieks and Vermaas eds.

\vspace{5mm}
\noindent
Barrett, J.\ A.: 1999 The Quantum Mechanics of Minds and Worlds,
Oxford University Press.

\vspace{5mm}
1996, `Empirical Adequacy and the Availability of Reliable
Records in Quantum Mechanics,' {\em Philosophy of Science\/} 63,
49--64.

\vspace{5mm}
1995, `The Distribution Postulate in Bohm's Theory,' {\em
Topoi\/} 14, 45--54.

\vspace{5mm}
\noindent
Bell, J.\ S.: 1987, {\em Speakable and Unspeakable in Quantum
Theory}, Cambridge University Press, Cambridge.

\vspace{5mm}
1982, `On the Impossible Pilot Wave,' {\em Foundations of
Physics\/} 12:989-899.  Reprinted in Bell (1987,159--168).

\vspace{5mm}
1981, `Quantum Mechanics for Cosmologists,' in {\em Quantum
Gravity\/} 2, C.\ Isham, R.\ Penrose, and D.\ Sciama (eds.),
Oxford: Clarendon Press, 611--637.  Reprinted in Bell
(1987,117--138).

\vspace{5mm}
1980, `de Broglie-Bohm, Delayed-Choice Double-Slit Experiment, and
Density Matrix,' {\em International Journal of Quantum Chemistry}:
Quantum Chemistry Symposium 14, 155--9.  Reprinted in Bell (1987, 111--6).

\vspace{5mm}
1976b, `The Measurement Theory of Everett and de Broglie's Pilot
wave,' in {\em Quantum Mechanics, Determinism, Causality, and
Particles}, M.\ Flato et al. (eds.), D. Reidel, Dordrecht,
Holland, 11--17.  Reprinted in Bell (1987, 93--99).

\vspace{5mm}
\noindent
Berndl, K., M.\ Daumer, D.\ D\"urr, S.\ Goldstein, and N.\
Zangh\'i: 1995, ``A Survey of Bohmian Mechanics,'' {\em Il Nuovo
Cimento}, vol.\ 110B, n.\ 5--6, 737--750.

\vspace{5mm}
Bohm, D.: 1952, `A Suggested Interpretation of Quantum Theory in Terms of
``Hidden Variables'', ' Parts I and II, {\em Physical Review\/}
85, 166--179, 180--193.

\vspace{5mm}
\noindent
Bohm, D.\ and B.\ J.\ Hiley: 1993, {\em The Undivided Universe:
An Ontological Interpretation of Quantum Theory}. London:
Routledge.

\vspace{5mm}
\noindent
Cushing, J.\ T.: 1996, `What Measurement Problem?' in R.\ Clifton
(ed.) (1996).

\vspace{5mm}
1994, {\em Quantum Mechanics: Historical Contingency and the
Copenhagan Hegemony}. Chicago: University of Chicago Press.

\vspace{5mm}
\noindent
Cushing, J.\ T., A.\ Fine, and S.\ Goldstein: 1996, {\em Bohmian Mechanics
and Quantum Theory: An Appraisal}, Boston Studies in the Philosophy of Science,
vol.\ 184, Kluwer Academic Publishers, Dordrecht, The Netherlands.

\vspace{5mm}
\noindent
Dewdney, C., L.\ Hardy, and E.\ J.\ Squires: 1993, `How Late
Measurements of Quantum Trajectories Can Fool a Detector,' {\em
Physics Letters A\/} 184, 6--11.

\vspace{5mm}
\noindent
Dickson, M.: 1996, ``Is the Bohm Theory Local?'', in J.\ T.\ Cushing et al.\ (eds) {\em Bohmian Mechanics and
Quantum Theory: An Appraisal}, (1996, 321--30).

\vspace{5mm}
\noindent
Dickson, M., R.\ Clifton: 1998, ``Lorentz-Invariance in Modal
Interpretations,'' forthcoming in Dieks and Vermaas (eds) (1998).

\vspace{5mm}
\noindent
Dieks, D.\ G.\ B.\ J., and P.\ E.\ Vermaas (eds): 1998, {\em The
Modal Interpretation of Quantum Mechanics}, Kluwer Academic
Press, forthcoming.

\vspace{5mm}
\noindent
Dirac, P.\ A.\ M.: 1958, {\em The Principles of Quantum Mechanics}, fourth edition,
Clarendon Press, Oxford.

\vspace{5mm}
\noindent
D\"urr, W.\ Fussender, S.\ Goldstein, and N. Zangh\`i: 1993,
``Comment on `Surrealistic Bohm Trajectories', '' {\em
Zeitschrift f\"ur Naturforschung\/} 48a, 1261--1262.

\vspace{5mm}
\noindent
D\"urr, D., S. Goldstein, and N. Zangh\'i: 1993, ``A Global
Equilibrium as the Foundation of Quantum Randomness,'' {\em
Foundations of Physics\/} 23, no.\ 5, 721--738.

\vspace{5mm}
1992, `Quantum Mechanics, Randomness, and Deterministic Reality',
{\em Physics Letters A\/} 172, 6-12.

\vspace{5mm}
1992, ``Quantum Equilibrium and the Origin of Absolute
Uncertainty,'' {\em Journal of Statistical Physics}, vol.\ 67,
nos. 5--6, 843--907.

\vspace{5mm}
\noindent
Englert, B.\ G., M.\ O.\ Scully, G.\ S\"ussmann, and H.\ Walther:
1993, ``Reply to Comment on `Surreal Bohm Trajectories', ''{\em
Zeitschrift f\"ur Naturforschung\/} 48a, 1263--1264.

\vspace{5mm}
1992, ``Surrealistic Bohm Trajectories,'' {\em Zeitschrift f\"ur
Naturforschung\/} 47a, 1175--1186.

\vspace{5mm}
\noindent
Maudlin, T.: 1994, {\em Quantum Nonlocality and Relativity}, Oxford,
Blackwell.

\vspace{5mm}
\noindent
Phillipidas, C.\, C.\ Dewdney, and B.\ H.\ Hiley: 1979, `Quantum Interference
and the Quantum Potential', {\em Il Nuovo Cimento\/} 52B, pp.\ 15--28.

\vspace{5mm}
\noindent
von Neumann, J.: 1955, {\em Mathematical Foundations of Quantum
Mechanics}, Princeton University Press, Princeton; translated by
R.\  Beyer from {\em Mathematische Grundlagen der
Quantenmechanik}, Springer, Berlin, 1932.

\end{document}